# High-pressure synthesis of $K_4N_6$ compound entirely composed of aromatic hexazine $[N_6]^{4-}$ anion

Jie Zhang[1,#], Tingting Ye[1,#], Guo Chen[1,2], Deyuan Yao[1,2], Xin Zhang[1,2], Junfeng Ding[1,2], Xianlong Wang[1,2,*]

[1]Key Laboratory of Materials Physics, Institute of Solid State Physics, Chinese Academy of Sciences, Hefei 230031, China

[2]University of Science and Technology of China, Hefei 230026, China

[#]These authors contributed equally to this work.

[*]Corresponding author

Email: xlwang@theory.issp.ac.cn

**Abstract**

The synthesis of hexazine $N_6$ ring is another milestone in nitrogen chemistry after that of aromatic $[N_5]^-$ anion. However, due to the diversity of carried charges, realizing compounds entirely composed of aromatic hexazine $N_6$ ring potentially with high-stability is a challenge. The first reported hexazine $N_6$ ring is $[N_6]^{2-}$ anion in $K_2N_6$ [Nat. Chem. 14, 794 (2022)] that does not adhere to Hückel's rule, and subsequently, the aromatic hexazine $[N_6]^{4-}$ anion mixed with $[N_5]^-$ anion and $N_2$ dimers is realized in the complex compound $K_9N_{56}$ [Nat. Chem. 15, 641 (2023)], where 5.36% of N atoms form aromatic $N_6$ ring. Here, we theoretically predict that all N atoms form aromatic hexazine $[N_6]^{4-}$ anion in $K_4N_6$, which becomes stable at 60 GPa and can stably exist up to 600 K at 0 GPa. Following this approach, based on the diamond anvil cell, $K_4N_6$ composed of 100% aromatic hexazine $[N_6]^{4-}$ anion is synthesized at 45 GPa after laser-heating and identified by synchrotron X-ray diffraction and Raman spectroscopy. Our results bring us closer to achieving aromatic $N_6$ rings at ambient condition.

**Introduction**

Polynitrogen molecules composed of only nitrogen atoms have long been a subject of interest among researchers in the field of high-energy density materials due to their favorable detonation properties. Despite extensive theoretical studies on numerous polynitrogen molecules [1, 2, 3, 4], their synthesis still poses significant challenges. Currently, only a few polynitrogen molecules have been identified in the condensed phase, including $N_3^-$, $N_5^+$, $N_5^-$, $N_6^{2-}$ and $N_6^{4-}$ [5, 6, 7, 8, 9, 10, 11, 12, 13]. Among these species, the $[N_5]^-$ anion is the first synthesized nitrogen ring, while the $N_6$ ring stands out as having the highest number of nitrogen atom. However, $N_6$ ring is more complex due to the existence of different forms, and its study has a rich historical background [2, 14, 15, 16], confirming its importance in nitrogen chemistry.

The $N_6$ ring was first experimentally detected by low-temperature photolysis in 1980 [14], which further motivates ongoing research efforts. Exactly, high pressures and temperatures provide conditions for eliminating N≡N triple bond, thereby making the synthesis of polymeric nitrogen feasible. Until recent years, successful synthesis of the elusive $N_6$ rings was achieved by laser heating in diamond anvil cell (DAC), which exhibit distinct characteristics based on their charge or configuration. The armchair-like $N_6$ ring in $WN_6$ was synthesized at ~126 GPa [17], followed by the synthesis of the planar $N_6$ ring with two electrons in the compound $K_2N_6$ at ~45 GPa [12]. Subsequently, the aromatic $N_6$ ring is synthesized in $K_9N_{56}$ at 46 GPa [13]. Different from the $N_6$ ring in $K_2N_6$, the planar $N_6$ ring in $K_9N_{56}$ carries four electrons, which adheres to Hückel's rule [18], suggesting the aromatic characteristic with enhancing stability. However, in the synthesized compound $K_9N_{56}$, 5.36% of nitrogen atoms form $N_6$ rings, while others form $N_2$ dimers and $N_5$ rings, leading to a lower content of $N_6$ ring [13]. In view of high stability of aromatic hexazine $[N_6]^{4-}$, the compounds with all nitrogen atoms forming aromatic $[N_6]^{4-}$ rings are desired.

In this work, the crystal structures of binary K-N system are searched. Especially, a $K_4N_6$ compound with all of the nitrogen atoms forming aromatic $N_6$ ring is identified. The experiment is designed to synthesize the $K_4N_6$ compound, where we try to react $KN_3$ and K by laser heating in the DAC. Finally, the desired $K_4N_6$ containing targeted hexazine $[N_6]^{4-}$ rings is synthesized at 45 GPa and 2000 K.

**Results and discussions**

### Theoretical prediction and properties of $K_4N_6$

After extensive structural searches in the K-N system under high-pressure, several stable compositions are identified, as detailed in the supplementary Fig S1. We find that $K_4N_6$ featuring $N_6$ rings is located on the convex hull at 60 GPa. The stability range of $K_4N_6$ is determined as 60-100 GPa according to its formation enthalpy, with the adoption of a $P\bar{1}$ structure as illustrated in Fig. 1a, a configuration also reported in previous studies [19]. Notably, $K_4N_6$ contains two additional K atoms compared to the previously reported $K_2N_6$ compound (Fig. 1b) [12, 20], suggesting the $N_6$ ring in $K_4N_6$ can carry a higher charge.

The electronic structure and density of states are depicted in Fig. 1c, showcasing $K_4N_6$ as an insulator with a band gap of 0.96 eV at 0 GPa. Near the fermi level, the density of states is mainly contributed by the $2p$ orbital electron of nitrogen, while the attribution of potassium is not evident, implying the loss of electrons by potassium atoms. Bader charge analysis [21] confirms that each potassium atom in the $K_4N_6$ compound losses 0.71 $e$ at 60 GPa. Notably, the $N_6$ ring in $K_4N_6$ has about twice the charge of the $N_6$ ring in $K_2N_6$, suggesting the formation of $[N_6]^{4-}$ ion in the $K_4N_6$ phase. Thus, the $N_6$ ring in $K_4N_6$ exhibits aromaticity with 10 π electrons, which satisfies the 4n+2 criteria. Additionally, the charge density exhibits a characteristic of delocalization (Fig. 1d), further substantiating the aromatic nature of the $N_6$ ring in $K_4N_6$. While the existence of $[N_6]^{4-}$ was previously noted in the $K_9N_{56}$ compound, its proportion represented 5.36% of the total nitrogen content [13]. In stark contrast, in the $K_4N_6$ compound, every nitrogen atom forms part of an aromatic $N_6$ ring, emphasizing the uniqueness and purity of the aromatic $N_6$ ring composition.

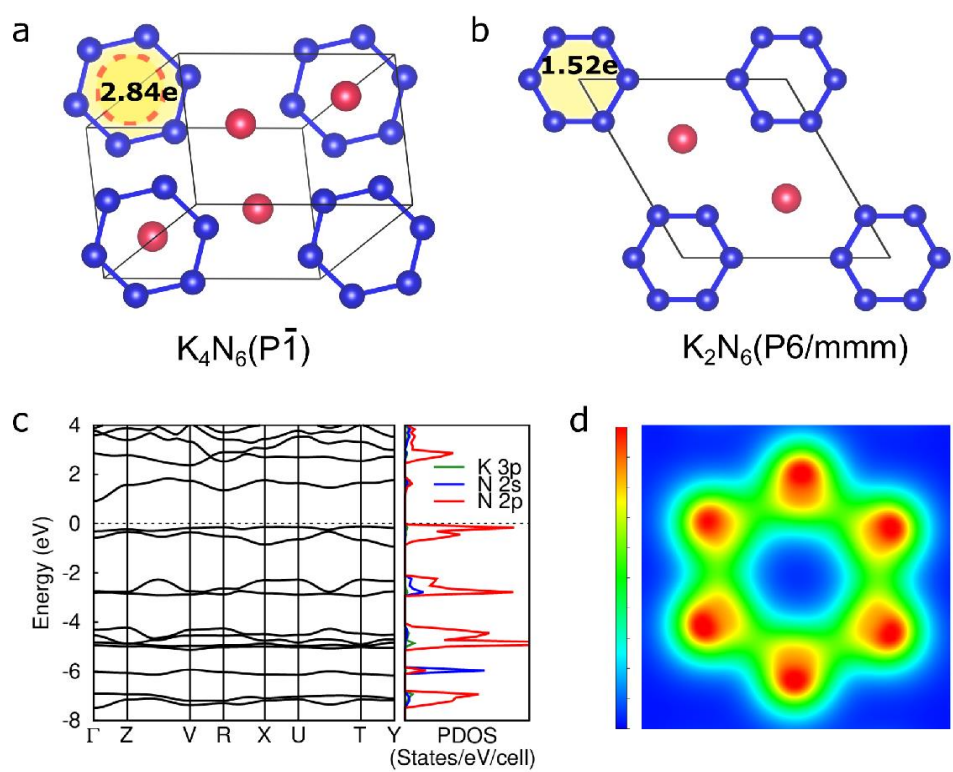

Fig. 1 **a**, **b**, Crystal structures of $K_4N_6$ and $K_2N_6$ compounds. Both the unit cell of $K_4N_6$ with $P\bar{1}$ symmetry and $K_2N_6$ with P6/mmm symmetry contain one $N_6$ ring. The values on the structures indicate the charge carried by each $N_6$ ring. The blue and red spheres represent N and K atoms, respectively. **c**, Electronic band structure and projected density of states (PDOS) for $P\bar{1}$ phase of $K_4N_6$ at 0 GPa. **d**, The charge density of the $N_6$ ring in $K_4N_6$ at 60 GPa. The isosurface value is 0.6. The electron distribution is nearly uniform, aligning with the characteristic of aromaticity.

The dynamical stability of the $K_4N_6$ structure is examined by calculating the phonon spectra. The absence of imaginary frequencies at 100 GPa (Fig. S2) signifies the dynamical stability of this structures, which is further confirmed upon releasing the pressure back to 0 GPa (Fig. 2a). Furthermore, first-principles molecular dynamical simulations are performed by using an NPT ensemble at 0 GPa. The root mean squared displacement (RMSD) of nitrogen atoms depicted in Fig. 2b reveals oscillates around a mean value below 700 K. As the temperature increases from 300 K to 700 K, the mean RMSD values are 0.427, 0.524, 0.549, 0.616 and 0.628 Å, indicating minimal thermal vibration of the nitrogen atoms in this temperature interval. While at 800 K, the mean

RMSD value is 1.688 Å and displays a near-linear dependence on time, suggesting atomic diffusion in the $K_4N_6$ compound. As shown in Fig. 2 c-e, $N_6$ rings start to decompose into $N_2$ dimers at 700 K, with a significant amount of $N_2$ dimers forming at 800 K. However, the $N_6$ rings are remained at 600 K, indicating the $K_4N_6$ compound with aromatic $N_6$ rings is stable at this temperature. These findings indicate that the $K_4N_6$ compound may be quenchable to the ambient conditions, emphasizing its potential for practical applications. Furthermore, inspired by the anticipated presence of the aromatic $N_6$ rings in the $K_4N_6$ phase, corresponding experimental investigations have been designed and conducted.

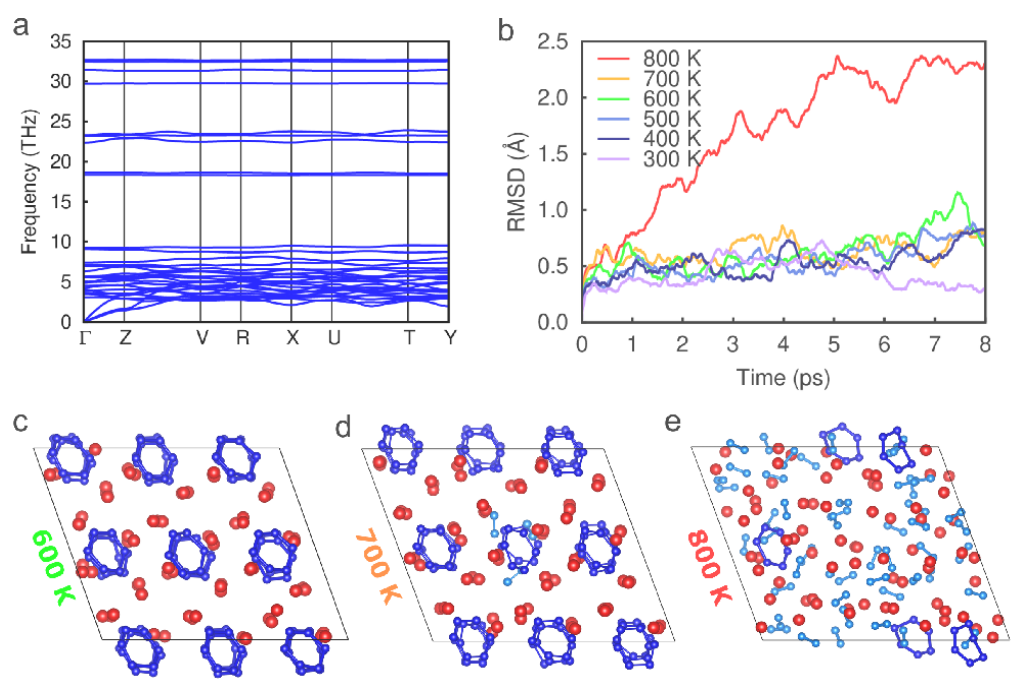

Fig. 2 **a**, Phonon spectra of $P\bar{1}$-$K_4N_6$ at 0 GPa. The disappearance of imaginary frequency of phonon spectra confirms the dynamical stability of $P\bar{1}$-$K_4N_6$. **b**, Root mean squared displacement (RMSD) of $P\bar{1}$-$K_4N_6$ at 0 GPa. RMSD of N atoms at the temperature of 300-800 K. The steep increase in RMSD at 800 K suggests that most $N_6$ rings are breaking down into $N_2$ dimers. **c**, **d**, and **e**, Structures of the $K_4N_6$ after molecular dynamics simulations at 600 K, 700 K and 800 K. The N atoms in $N_6$ rings and $N_2$ dimers are represented by the blue and light blue spheres, respectively, while K atoms are shown as the red spheres.

### Synthesis of $K_4N_6$ at high-pressure and high-temperature

Regarding the formation of $K_2N_6$ from pure $KN_3$ and the preparation of $K_9N_{56}$ from $KN_3$ mixed with $N_2$ at high-pressure and high-temperature, the elemental ratio 4:6 of K:N suggests that a potassium-rich environment is required to synthesize the $K_4N_6$ compound. Thus, we compressed powdered $KN_3$ and micrometer-size pure K pieces without pressure transmitting medium/reagent in DACs (Fig. 3a). To facilitate a chemical reaction between $KN_3$ and K, samples were then laser-heated using YAG lasers at a pressure range of 40-50 GPa, and it is worth noting that the heated spot should contain both $KN_3$ and K. We found that laser-heating at least 2000 K at a pressure range of 45-50 GPa resulted in the formation of the target compound, $K_4N_6$ (space group $P\bar{1}$), identified by both Raman and X-ray diffraction spectroscopy measurements.

In Raman measurement, multiple phases were identified as coexisting in the DAC, including $KN_3$, $K_2N_6$, and a new phase with a set of Raman modes emerging from the heated spot that do not correspond to known phases of $KN_3$, $K_2N_6$, and $K_9N_{56}$ (Fig. 3 b). Firstly, compare with the Raman peaks of pure $KN_3$ at 50 GPa, the disappearance of intense mode characteristic of the $N_3$ azide anion indicates the chemical reaction of $KN_3$ after laser-heating, and the broad lattice bands observed in the spectral range of 100-600 cm$^{-1}$ transformed to sharp pronounced Raman peaks at 162, 210, 250, 313, 360, 407, and 534 cm$^{-1}$, corresponding to the N-N bending modes and lattice vibrations, thereby clearly indicates the synthesis of a new K-N compound on the heated spot in this work. Secondly, the Raman peaks of the new compound in the range of 1000-1300 cm$^{-1}$ denotes the signal of $N_6$ ring as reported in the earlier works of K-N compounds [12]. In contrast to the Raman spectrum of $K_2N_6$ at 50 GPa and $K_9N_{56}$ in literature, the low-frequency Raman bands form 0-500 cm$^{-1}$ are significantly different, which shows the formation of a new K-N compound containing $N_6$ ring on the heated spot. Thirdly, based on the theoretical predicted structure of $K_4N_6$, the corresponding Raman frequencies were calculated and compared with observed Raman peak. All the peaks could be well assigned to the Raman modes of $K_4N_6$. The good agreement between theoretical calculation and experimental observation suggests the synthesis of $K_4N_6$ with $N_6$ ring.

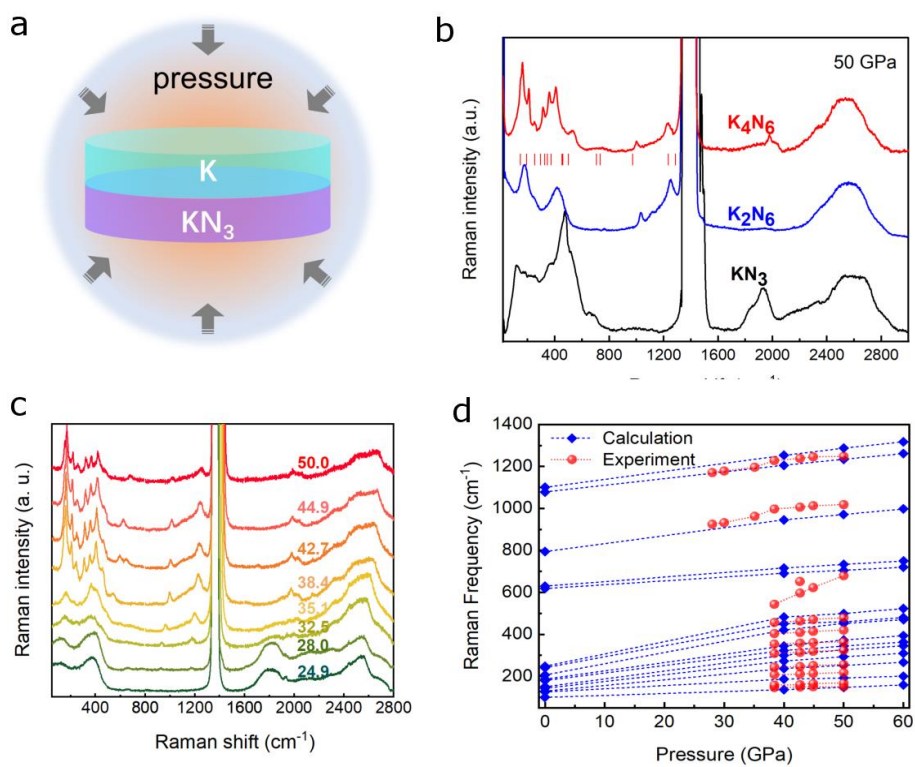

**Fig. 3 a**, Schematic of KN$_3$ and K under pressure. **b**, Raman spectra of and K$_4$N$_6$, K$_2$N$_6$, and KN$_3$ around 50 GPa. The ticks show the theoretically calculated positions of the Raman modes of K$_4$N$_6$ with P$\bar{1}$ symmetry. **c**, Evolution of the Raman modes of the K$_4$N$_6$ compounds upon pressure release at room temperature. **d**, The calculated and experimental Raman frequencies as a function of pressure.

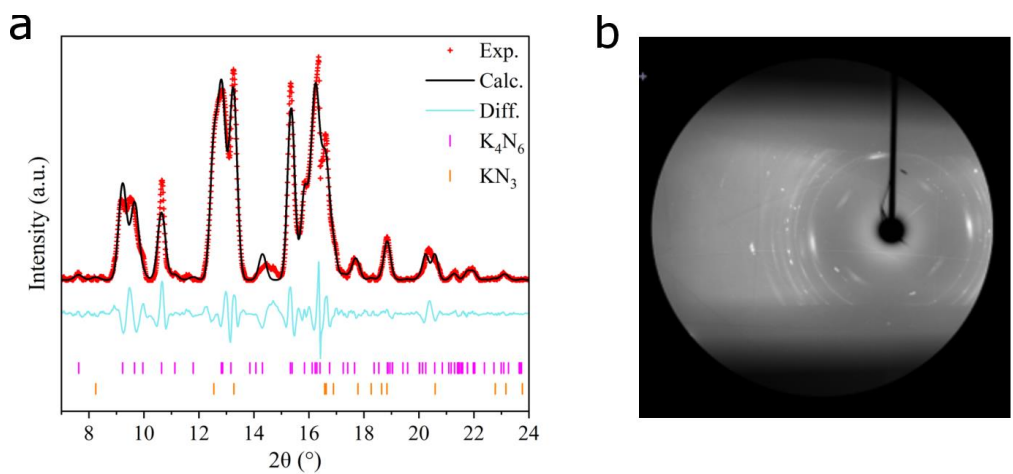

**Fig4 a**, Powder-crystal X-ray diffraction data in compressed to 50 GPa and laser-heated via direct coupling KN$_3$ and K. **b**, Raw diffraction images of the sample.

The frequencies of the characteristic N-N stretching deformation modes decrease with pressure, and the sharp low-frequency Raman peaks gradually weaken as the pressure decreases, finally disappearing at 28 GPa (Fig. 3c). Based on the disappearance of the Raman spectrum corresponding to the stretching vibration of nitrogen observed in the spectral range of 1000-1300 cm$^{-1}$, we have been able to trace the Raman spectrum of this compound down to approximately 28 GPa upon pressure release. Thus, their pressure behavior is established and compared with theoretical calculations of vibrational frequencies of $P\bar{1}$ $K_4N_6$ at 0-60 GPa (Fig. 3d). Although the number of the calculated Raman modes exceeds the measured ones, their values match closely. The pressure dependencies of the fundamental modes of the $[N_6]^{4-}$ ring in $K_4N_6$ exhibit trends similar to those of $[N_6]^{2-}$ in $K_2N_6$ [12] and $[N_5]^-$ in $NaN_5$ [9].

To identify the crystal structure of the synthesized $K_4N_6$ compound, in-situ high pressure synchrotron X-ray diffraction experiments were carried out, and the structure refinement against the diffraction data of the temperature-quenched sample is performed. The detailed comparison between the theoretically predicted model of $K_4N_6$ and the experimental spectra reveal the phase of the $K_4N_6$ compound as our predicted $K_4N_6$ with $P\bar{1}$ space group (Fig. 4). The lattice parameters at 50 GPa is detailed in the Table S1 of supplementary materials. The refined XRD pattern supports the successfully synthesized compound $K_4N_6$ with all of nitrogen atoms forming aromatic $N_6$ rings.

Through a combination of theoretical and experimental research, we successfully synthesized a new compound called $K_4N_6$ at high pressure. The potassium-rich environment provided the necessary electrons for the formation of $[N_6]^{4-}$ ring, highlighting the significance of charge transfer in the synthesis process. However, the aromatic $[N_6]^{4-}$ is not preserved at the lower pressure, similar to decomposition pressure with the $[N_6]^{2-}$. This may be due to changes in the DAC environment caused by the decomposition of $[N_6]^{2-}$, which destabilizes the aromatic $[N_6]^{4-}$. In the future, the stability may be enhanced by synthesizing more $K_4N_6$ samples to avoid the influence of nonaromatic $[N_6]^{2-}$, with the goal of capturing the aromatic $N_6$ ring at ambient pressure. Moreover, based on insights gleaned from the examination of the $N_5$ ring and cg-N [7,10,11,22,23,24,25,26], we expect that the aromatic $N_6$ rings in $K_4N_6$ will be achievable under ambient conditions in future research.

## Conclusions

A predicted compound known as $K_4N_6$, exclusively composed of aromatic $[N_6]^{4-}$ rings, is synthesized through the reaction between $KN_3$ and K at a pressure of 45 GPa and a temperature of 2000 K. By establishing a K-rich environment, there are sufficient electrons available to transfer to the nitrogen atoms. This process results in a pressure-induced transition from $N_3$ azides to $N_6$ rings, with each $N_6$ ring holding four electrons within the $K_4N_6$ compound. Notably, every nitrogen atom actively contributes to the configuration of $N_6$ rings in this distinctive chemical structure, elucidating an innovative pathway for producing aromatic $N_6$ rings.